# Project Pipeline: Preservation, Persistence, and Performance

*Enabling HIVE for Accessing and Using Historical Vocabularies*

**Jane Greenberg**
Drexel University
USA
jg3243@drexel.edu
0000-0001-7819-5360

**Christopher B. Rauch**
Drexel University
USA
cr625@drexel.edu
0000-0003-2061-3413

**Mat Kelly**
Drexel University
USA
mkelly@drexel.edu
0000-0002-0236-7389

**Abstract** – Preservation pipelines demonstrate extended value when digitized content is also computation-ready. Expanding this to historical controlled vocabularies published in analog format requires additional steps if they are to be fully leveraged for research. This paper reports on work addressing this challenge. We report on a pipeline and project progress addressing three key goals: 1) transforming the 1910 Library of Congress Subject Headings (LCSH) to the Simple Knowledge Organization System (SKOS) linked data standard, 2) implementing persistent identifiers (PIDs) and launching our prototype ARK resolver, and 3) importing the 1910 LCSH into the Helping Interdisciplinary Vocabulary Engineering (HIVE) System to support automatic metadata generation and scholarly analysis of the historical record. The discussion considers the implications of our work in the broader context of preservation, and the conclusion summarizes our work and identifies next steps.

**Conference Topics** – Exploring the New Horizons

## I. Introduction

Technological advances have transformed our methods and processes for preserving and accessing the world of recorded knowledge. Take, for example, the digitization of the Dead Sea Scrolls–a historical record of great cultural, linguistic, and religious significance. Technological innovation has supported both the conservation of the original, artifactual fragments as well as the production of high-resolution digital facsimiles, enabling worldwide access to the intellectual content of these coveted texts. The digital renderings of these historical records have enabled the application of AI methods, specifically digital paleography techniques such as feature extraction, handwriting, and pattern recognition [1].

Preservation of the Dead Sea Scrolls has also included transforming the writings to digital text via optical character recognition (OCR), which has then been translated to several other languages. These computation-ready textual renderings of the Dead Sea Scrolls, as seen with other historical records, present a new horizon for study [2],[3],[4].

As exciting as these developments are, this paradigm shift cannot be complete without transforming *historical vocabularies* reflective of the content contained within these resources into a computational format. This need has been a growing focus of the Metadata Research Center at Drexel University, where a team of researchers and affiliated partners are advancing the pipeline supporting the *preservation, persistence, and performance* of historical terminologies.

This paper presents our work addressing three key goals. The first goal has been to enhance the preservation of historical terminologies by transforming their digital renderings into computation-ready text. As part of this phase, we adopt the Simple Knowledge Organization System (SKOS) linked data standards, to enable computation. The second goal, in line with both preservation and linked data best practices, has been the implementation of persistent identifiers (PIDs). Our initial case study is the 1910 Library of Congress Subject Headings [5],[6], with new cases underway. The third goal of this effort has been to enable computation, hence the performance of the transformed terminology in the Helping Interdisciplinary Vocabulary Engineering (HIVE) Systems. The paper reports on these three goals, discusses current implications, and identifies next steps.

## II. Background/Literature Review

### A. Digital Preservation and Computation

Digital preservation as a focused topic of study commenced in the 1990s, motivated largely by the launch of the World Wide Web and the enhanced capability for sharing digital content. The web accelerated challenges associated with versioning by bringing questions associated with early computing in the 1970s to a much wider and more diverse group of constituents in infor-mation and library science, computing, and intersecting disciplines. The launch of the Internet Archive in 1996 [7], publication of *Preserving Digital Information: Report of the Task Force on Archiving of Digital Information* [8], and initiation of the National Digital Information Infrastructure and Preservation Program (NDIIPP) program [9], helped to spearhead and advance the U.S. national preservation strategy. Globally, there have been equally important developments, such as the initial call from the Chinese Academy of Science (CAS), hosting the International Conference on Digital Preservation (iPRES) in Beijing [10], and the work of the eIFL (Electronic Information for Libraries) [11], as well as the European Commission Recommendation on digitising and digital preservation [12].

Preservation advances have brought to the forefront the goal of making digital resources computation-ready. This goal is part of two key actions embedded in the outer rim of Digital Curation Lifecycle Model [13]:

**Action 1: Access, Use and Reuse**
Ensure that data is accessible to both designated users and reusers, on a day-to-day basis. This may be in the form of publicly available published information. Robust access controls and authentication procedures may be applicable.

**Action 2: Transform**
Create new data from the original, for example, by migration into a different format or by creating a subset, by selection or query, to create newly derived results, perhaps for publication.

*B. Historical Vocabularies and Persistence*

Researchers have recognized that contemporary vocabularies (e.g., controlled vocabularies, ontologies, taxonomies, and related systems) fail to sufficiently capture the relevant contextual knowledge required for search, retrieval, and analysis of historical documents [14]. A vocabulary temporally aligned with the resource better reflects how knowledge was understood at the time [15]. Further, the temporal alignment between vocabulary and resources may aid machine learning and enrich computational analysis of historical text.

In pursuing historical vocabularies, researchers must also be mindful of biases, offensive language, and other contemporary problematic features of the past and develop systems that enable them to track changes. The attribution of persistent identifiers to both historical and contemporary terms provides a means for ascribing these associations and assists in attaining the goal of making historical vocabularies accessible for research. This potential applicability of PIDs to historical vocabularies has also motivated us to address the persistence of these vocabularies.

Persistent identifiers are foundational elements in the building of information and research infrastructure [16]. Globally or contextually unique identifiers provide a way to refer to an information resource unambiguously. No single institution or archive holds a complete record of an information resource [17]. The preservation of digital resources is technically complex and for preservation, multifaceted objects must necessarily be fragmented into a number of single objects [18]. In addition to comprising the components of the amalgamated description of a preserved item, related data can provide important contextual data to facilitate automatic processes such as entity extraction and text summarization.

Persistence and preservation go hand in hand – persistence implies longevity. The longevity of standardized terminologies is reinforced with the implementation of a persistent identifier system that is sufficiently flexible and granular enough to track changes over time. There are several fairly well-recognized PID formats that are applied to scholarly resources. Examples include digital object identifiers (DOIs), International Standard Book Numbers (ISBNs), International Standard Serial Numbers (ISSNs), and Archival Resource Keys (ARKs). A key requirement for a persistent identifier system to be applied to historical vocabularies is to support each published release of an authorized term over time, enabling unique identification. This is critical for supporting a predictable, effective, reference structure that enables navigation of the terms' lifecycle. As a result, the PID system, in combination with a queryable data source, will enable automatic linking and analysis on a larger scale.

Archival Resource Keys (ARKs) support these needs with a decentralized infrastructure. ARKs are intended to serve as stable, portable, trusted unique references to any category of resource, whether digital, physical, or abstract. The ARK format also supports the notion of inflection, which allows qualifiers to be appended to the identifiers in which the ARK is embedded. For example, the URI encapsulating an ARK may have a qualifier that acts as an API key to return the SKOS RDF description of an object, data in JSON or XML format, filtered data, or even raw image data. We have taken initial steps in this direction, developing the foundation for an API gateway into the vocabulary server application HIVE (Helping Interdisciplinary Vocabulary Engineering). We report on this work and connected developments in Section III.

III. Project Pipeline: Preservation, Persistence, and Performance

Over the past year, we have progressed toward our goals of enhancing the preservation of historical vocabularies by making them computational, implementing persistent identifiers (PIDs), and enabling computational use of historical vocabularies. In this section, we report on our efforts of preservation (Section III A), (Section III B), and performance (Section III C).

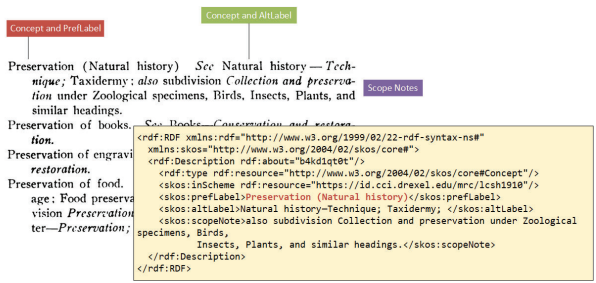

Figure 1: SKOS-enabled vocabularies representing the Library of Congress 1910 Subject Headings

### A. Preservation: Transforming a Historical Vocabulary into a Computational Format

The first goal in our work plan is to enhance the preservation aspect by transforming a historical vocabulary into a computation-ready rendering. Our use case involves the Library of Congress Subject Headings (LCSH). This is one of the most well known general domain controlled vocabularies used for creating standardized topical metadata to describe analog as well as digital scholarly resources. Historical versions of the LCSH have been digitized and made available in open repositories. The HaithiTrust is one such repository [1], offering access to high-quality PDF versions of previous editions dating back to the LCSH 1910. The 1910 LCSH was selected for our initial case study, interconnected with the National Endowment for the Humanities (NEH) supported 19th Century Knowledge Project, a collaboration involving the Metadata Research Center and Temple University Libraries. [2]

The specific task of transforming the vocabulary involved converting the PDF version into an eXtensive Markup Language-Text Encoding Initiative (XML-TEI) file. Next, we developed a set of rules to guide the conversion of the XML-TEI into the Simple Knowledge Organization System (SKOS) format. The rules we developed aligned with current thesauri standards [19], where applicable. For example, the 1910 LCSH's use of *See*, indicating "See from A to B" where B refers to the authorized term (SKOS: PrefLabel). Additionally, the term preceding a See reference stands (A) is less favorable, but synonymous, and in SKOS is rendered as an AltLabel. Figure 1 captures how terms appear in a digital facsimile of the 1910 LCSH (left-hand side), and the SKOS encoding (right-hand side). The SKOS version on its own is robust and could be used in any information systems supporting SKOS-formatted data.

As discussed below, we added ARKs to enable persistence, and the vocabulary is imported into HIVE enabling the performance.

---

[1] https://www.hathitrust.org
[2] https://tu-plogan.github.io/source/c_introduction.html

### B. Persistence

The second goal in our work plan of persistence is twofold. We aimed to align with best practices surrounding persistence as it relates to preservation as well as the attribution of PIDs and enabling the potential for linked data. We adopted the ARK PID system (Section II B) for the first requirement, which exhibits a decentralized approach toward persistence and allows for decentralized dereferencing through embedding in URIs.

ARKs require a resolver to associate a resource with the assigned key. Some institutions maintain their own resolvers, while others utilize the Names To Things (N2T) service, supported by the California Digital Library. There are also hybrid approaches, as we implemented, to identify terms of the controlled vocabulary made available by the HIVE service. For example, the HIVE server contains the SKOS-enabled version of the 1910 LCSH available through a REST API. The server responds to queries with JSON-encoded data representing the vocabulary entry. The returned data is then formatted for presentation on a public web server and made available through a dereferenceable version of the ARK prepended with the Name Mapping Authority (in effect, the hostname).

According to the ARK specification, the ARK identifier is independent of a particular hostname and will take a form like ark:/99152/b4057cr7r to represent an information object. There are currently two ways to resolve this reference. The first is the centralized Names to Things service (https://n2t.net/) `https://n2t.net/ark:/99152/b47p8tc5z`. This is a common resolver that is available to be used for all ARKs. The second is `https://id.cci.drexel.edu/ark:/99152/b47p8tc5z`. The Name Mapping Authority represents the institutional resolver at Drexel that responds to incoming requests for a vocabulary term in ARK format by presenting the data returned from the JSON API call to HIVE.

### C. Performance

Our third goal is to enable performance, specifically actionable use and operation of the 1910 LCSH vocabulary. This goal was achieved by importing the 1910 LCSH into the Helping Interdisciplinary Vocabulary Engineering (HIVE) application. HIVE is a linked data, automatic metadata generation application that supports access and use of a suite of controlled vocabularies [20], [21]. HIVE2, the most current system, is Python-based and supports SKOS encoded vocabularies. HIVE enables user navigation of an individual vocabulary as well as search across multiple vocabularies. HIVE also supports automatic metadata generation by drawing terms from one or more vocabularies, depending on the project's need.

The automatic functionality for HIVE currently uses the RAKE algorithm [22]. The sequence involves converting a resource (e.g. a document in text, PDF, and HTML format) into plain text. The document text is normalized, stop words removed, and text is filtered by term frequency, word length, and phrase length. The filtered text is then used to generate a list of keywords and phrases, which is matched against the terms in the selected vocabulary. An example of HIVE's output is shown in Figure 2.

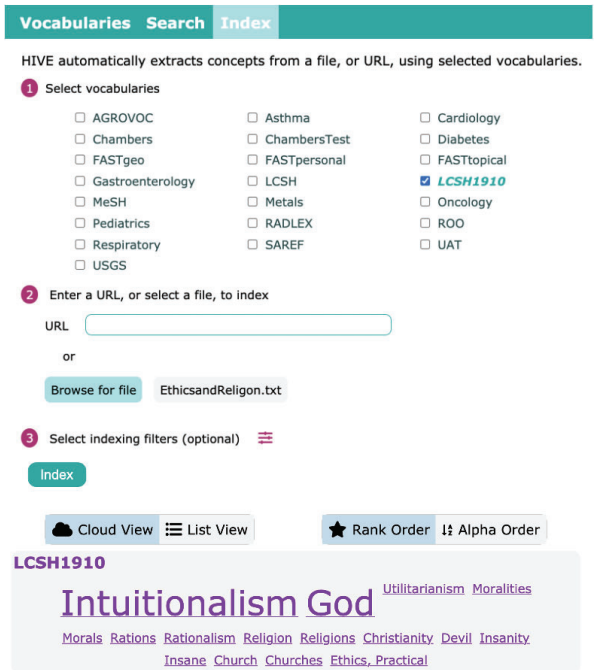

Figure 2: HIVE Tag Cloud Output

IV. Discussion

The work reported in this paper serves as a proof of concept for a pipeline that ultimately supports access and use of historical ontologies. The outcomes also give insight into how we might pursue future preservation-connected work. In considering the broader impact, we note here three particular insights to help serve as the basis for further exploration.

First, digital preservation efforts toward making text computation-ready open up new pathways. This is not a novel insight–our work with historical terminologies involved additional steps transforming text into SKOS. This type of opportunity may be integrated into preservation strategies, given that there are likely hundreds of historical vocabularies that are part of the historical record.

Second, persistence is a necessary component of preservation that is already embraced by the preservation community. PIDs support linking the history and evolution of a record over time. Related work with the Memento project [23] stands as a model, and potential future basis for further integrating temporally-pinned resources with the type of linked data discussed in this paper.

Finally, preservation initiatives continue to invite tool development. The development of repository and digital library software is interlinked with preservation . As demonstrated by HIVE, having tools that make preserved documents actionable can enhance the use and study of resources.

V. Conclusion

Historical documents are increasingly being transformed to digital formats and even text-enabling analytics, although temporally aligned vocabularies have not been addressed on the same level. This deficit presents a preservation and access opportunity, specifically transforming historical vocabularies from their printed or physical object forms into electronic forms and prepared as linked data. Research involving historical texts can then be indexed and analyzed according to the principles of knowledge organization at the time of authorship showing topical representation during the same time period.

The work presented in this paper is addressing is advancing the access, use, and reuse of historical vocabularies. Through the application of standardized markup, specifically using the SKOS format, and by provisioning the HIVE vocabulary service, we have made the 1910 LCSH computation-ready. In future work, we intend to expand the offering of additional historical vocabularies to cover additional time periods and conduct comparative research relating to entity extraction and summarization using lists associated with specific historical timeframes.

Acknowledgments

The authors wish to acknowledge the contributions of Peter M. Logan, John Kunze, Joan P. Boone, and Sam Grabus to this work as well as funding from NEH Grant #HAA-261228-18